\documentclass[letterpaper, 10 pt, conference]{ieeeconf}  
\usepackage{amsmath, amssymb}
\usepackage{graphicx}
\usepackage{subfigure}
\usepackage[noadjust]{cite}
\newtheorem{theorem}{Theorem}
\newtheorem{remark}{Remark}
\usepackage{color}
\usepackage{amsmath, amssymb}
\usepackage{graphicx}
\usepackage{subfigure}
\usepackage[noadjust]{cite}
\usepackage{color}
\usepackage[noadjust]{cite}
\usepackage{color}
\usepackage{wrapfig}

\usepackage{hyperref}
\usepackage{mathrsfs}
\usepackage{cite}
\IEEEoverridecommandlockouts                              
\title{\LARGE \bf
Towards Electronics-based Emergency Control in Power Grids with High Renewable Penetration
}

\author{Thanh Long Vu,~\IEEEmembership{Member,~IEEE,} Spyros Chatzivasileiadis,~\IEEEmembership{Member,~IEEE,} and~Konstantin~Turitsyn,~\IEEEmembership{Member,~IEEE}
\thanks{Thanh Long Vu, Spyros Chatzivasileiadis, and Konstantin Turitsyn are with the Department of Mechanical Engineering, Massachusetts Institute of Technology, Cambridge, MA, 02139 USA, e-mail: \{longvu, chatziva, turitsyn\}@mit.edu.
}}

\begin{document}

\maketitle
\thispagestyle{empty}
\pagestyle{empty}

\maketitle
\begin{abstract}
Traditional emergency control schemes in power systems usually accompany with power interruption yielding severely economic damages to customers. This paper sketches the ideas of a viable alternative for traditional remedial controls for power grids with high penetration of renewables, in which the renewables are integrated with synchronverters to mimic the dynamics of conventional generators. In this novel emergency control scheme, the power electronics resources are exploited to control the inertia and damping
of the imitated generators in order to quickly compensate for the deviations caused by fault and thereby bound the fault-on dynamics and stabilize the power system under emergency situations. This  emergency control not only saves investments and operating costs for modern and future power systems, but also helps to offer seamless electricity service to customers. Simple numerical simulation will be used to illustrate the concept of this paper.
\end{abstract}

\maketitle

\section{Introduction}

The aging US power grid is approaching its physical limits with the high penetration of intermittent renewables, large volume of power storage and EVs, and ubiquitous presence of massive loads. As a result, the stressed system is especially vulnerable to extreme events. Currently, the stability of power grid under emergency situations is mainly based on remedial actions \cite{119276,6965423,6939093}, special protection systems (SPS) \cite{SPS,982194} and load shedding \cite{141798, 1461637, 6345547} to quickly rebalance power and hopefully stabilize the system. However, these emergency actions mainly rely on interrupting electricity service to customers. The unexpected load shedding are extremely harmful to customers since it may lead to enormously high economic damage. On the other hand, the protective devices are usually only effective for individual elements, but less effective in preventing the grid from collapse, and in many cases may split the grid into islands or lead to cascading failures \cite{Koch2010}. The underlying reason is the lack of coordination among protective devices and the difference in their timescales, which together make them incapable to maintain the grid stability in a whole. 

These issues on economic efficiency and system stability  call for a new generation of emergency controls, which can guarantee the system stability recovery and reduce
the damages to customers. In this paper, we aim to bring the emergency control problem to the attention of the control community, and hence sketch the ideas of a novel emergency control scheme for renewable-integrated power grids by exploiting the emerging power electronics resources. Remarkably, this emergency control scheme can reduce the needs in power interruptions due to load shedding, and maintain the transient stability of power grids under emergency situations. 

In particular, this paper will bring in the following novelties:  

\begin{itemize}
\item We model the power grids with high levels of renewable penetration, in which the renewable generators are integrated with the synchronverters \cite{Zhong:2011}. The synchronverters will control the renewable generators to imitate the dynamics of the conventional generators, which is described by the classical swing equations. In addition, the inertia and damping parameters of these mimic generators can be quickly adjusted in a wide range of values.
\item Exploiting such electronics resources, we formulate the emergency control problem in this renewable-integrated power system, which aims to maintain the transient stability of the system following a line tripping by appropriately tune the inertia and damping of the imitated generators.
\item We solve this emergency control by applying our recently introduced quadratic Lyapunov function-based transient stability certificate \cite{VuTuritsyn:2015TAC}. In particular, we present sufficient conditions on the inertia and damping of the imitated generator such that when applied to the fault-on dynamics, the fault-cleared state after a fixed clearing time still stays inside the region of attraction of the post-fault equilibrium point. Note-worthily, the sufficient conditions in many case can be formulated as a set of linear matrix inequalities (LMIs), which can be solved quickly by convex optimization enabling the fast response of the remedial actions. 
\end{itemize}

The paper is structured as follows. In Section \ref{sec.model} we
introduce the standard structure-preserving model of power
systems including conventional/renewable generators and dynamic loads. On top of this model, we formulate the emergency control problem of
power grids, which aims to exploit the integrated power electronics resources to maintain the transient stability of the system. In Section \ref{sec.certificates} we recall
the recently introduced quadratic Lyapunov functions-based  certificate for transient stability, which will then be
instrumental to designing emergency control in this paper. In Section \ref{sec.emergencycontrol}, we sketch the emergency control design, propose the procedure for
emergency control practice, and discuss ways to overcome the issues of computation and regulation delays. Section \ref{sec.simulations}
illustrates the effectiveness of the proposed scheme based on
numerical simulation on a simple power system and Section \ref{sec.conclusion} concludes the paper.

\section{Network Model and Emergency Control Problem}
\label{sec.model}

\subsection{Network Model}

Consider a power transmission grid including conventional generators, renewable generators, loads, and
transmission lines connecting them. We assume that all the renewables are integrated with synchronverter \cite{Zhong:2011}, which will control the renewables to mimic
the dynamics of conventional generator, and thus we call both conventional and renewable generators as generators. A generator has both internal
AC generator bus and load bus. A load only has load bus but no
generator bus. Generators and loads have their own dynamics influenced by the nonlinear AC power flows in the transmission lines. 
In this paper we consider the standard structure-preserving model to describe components and dynamics in
power systems \cite{bergen1981structure}. This model naturally
incorporates the dynamics of generators' rotor angle as well as response of
load power output to frequency deviation. Although it does not
model the dynamics of voltages in the system, in comparison to the classical swing equation with constant impedance loads, the structure of power grids is preserved in this model. 

Mathematically, the grid is described by an undirected graph
$\mathcal{G}(\mathcal{N},\mathcal{E}),$ where
$\mathcal{N}=\{1,2,\dots,|\mathcal{N}|\}$ is the set of buses and
$\mathcal{E} \subseteq \mathcal{N} \times \mathcal{N}$ is the set
of transmission lines connecting those buses. Here, $|A|$ denotes
the number of elements in the set $A.$ The sets of conventional/renewable generator buses
and load buses are denoted by $\mathcal{G}=\mathcal{G_C} \cup \mathcal{G_R}$ and $\mathcal{L}$ and
labeled as $\{1,..., |\mathcal{G}|\}$ and $\{|\mathcal{G}|+1,...,
|\mathcal{N}|\}.$ We assume that the grid is lossless with
constant voltage magnitudes $V_k, k\in \mathcal{N},$ and the
reactive powers are ignored.


\textbf{Conventional generator buses.} The dynamics of conventional generators are described
by a set of the so-called swing equations:
\begin{align}
\label{eq.swing1}
  m_k \ddot{\delta_k} + d_k \dot{\delta_k} + P_{e_k}-P_{m_k}
  =0, k \in \mathcal{G_C},
\end{align}
where, $m_k>0$ is the dimensionless moment of inertia of the
generator, $d_k>0$ is the term representing primary frequency
controller action on the governor. $P_{m_k}$ is the effective
dimensionless mechanical torque acting on the rotor and $P_{e_k}$
is the effective dimensionless electrical power output of the
$k^{th}$ generator.

\textbf{Synchronverter-integrated renewable generator buses.}
The dynamics of synchronverter-based renewable generators are described
by the same set of equations:
\begin{align}
\label{eq.swing1renewable}
  m_k \ddot{\delta_k} + d_k \dot{\delta_k} + P_{e_k}-P_{m_k}
  =0, k \in \mathcal{G_R},
\end{align}
where the inertia $m_k$ and damping $d_k$ are tunable.

\textbf{Load buses.} Let $P_{d_k}$ be the real power drawn by the
load at $k^{th}$ bus, $k \in \mathcal{L}$. In general $P_{d_k}$ is
a nonlinear function of voltage and frequency. For constant
voltages and small frequency variations around the operating point
$P^0_{d_k}$, it is reasonable to assume that
\begin{align}
P_{d_k}=P^0_{d_k} + d_k \dot{\delta}_k, k \in \mathcal{L},
\end{align}
where $d_k>0$ is  the constant frequency coefficient of load.

\textbf{AC power flows.} The active electrical power $P_{e_k}$
injected from the
$k^{th}$ bus into the network, where $k \in \mathcal{N},$ is given by
\begin{align}
\label{eq.electricpower}
  P_{e_k}=\sum_{j \in
  \mathcal{N}_k} V_kV_jB_{kj} \sin(\delta_k
  -\delta_j), k\in \mathcal{N}
\end{align}
Here, the value $V_k$ represents the voltage magnitude of the
$k^{th}$ bus which is assumed to be constant. $B_{kj}$ are the
(normalized)  susceptance of the transmission line $\{k,j\}$ connecting the $k^{th}$ bus and $j^{th}$ bus.
$\mathcal{N}_k$ is the set of neighboring buses of the $k^{th}$
bus. Let $a_{kj}=V_kV_jB_{kj}.$ 

By power balancing we obtain the
structure-preserving model of power systems as:
\begin{subequations}
\label{eq.structure-preserving}
\begin{align}
\label{eq.structure-preserving1}
 m_k \ddot{\delta_k} + d_k \dot{\delta_k} + \sum_{j \in
  \mathcal{N}_k} a_{kj} \sin(\delta_k-\delta_j) = &P_{m_k},  k \in \mathcal{G},  \\
  \label{eq.structure-preserving2}
  d_k \dot{\delta_k} + \sum_{j \in
  \mathcal{N}_k} a_{kj} \sin(\delta_k-\delta_j) = &-P^0_{d_k},  k \in \mathcal{L},
\end{align}
\end{subequations}
where, the equations \eqref{eq.structure-preserving1} represent
the dynamics at generator buses and the equations
\eqref{eq.structure-preserving2} the dynamics at  load buses.

The system described by equations \eqref{eq.structure-preserving}
has many stationary points with at least one stable corresponding
to the desired operating point. Mathematically, the state of \eqref{eq.structure-preserving} is presented by 
$\delta=[\delta_1,...,\delta_{|\mathcal{G}|},\dot{\delta}_1,...,\dot{\delta}_{|\mathcal{G}|},\delta_{|\mathcal{G}|+1},...,\delta_{|\mathcal{N}|}]^T,$ and the desired operating point is
characterized by the buses' angles
$\delta^*=[\delta_1^*,...,\delta_{|\mathcal{G}|}^*,0,\dots,0,\delta^*_{|\mathcal{G}|+1},...,\delta^*_{|\mathcal{N}|}]^T.$ This point
is not unique since any shift in the buses' angles
$[\delta_1^*+c,...,\delta_{|\mathcal{G}|}^*+c,0,\dots,0,\delta^*_{|\mathcal{G}|+1}+c,...,\delta^*_{|\mathcal{N}|}+c]^T$
is also an equilibrium. However, it is unambiguously characterized
by the angle differences $\delta_{kj}^*=\delta_k^*-\delta_j^*$
that solve the following system of power-flow like equations:
\begin{align}
  \label{eq.SEP}
  \sum_{j \in
  \mathcal{N}_k} a_{kj} \sin(\delta_{kj}^*) =P_{k}, k \in \mathcal{N},
\end{align}
where $P_k=P_{m_k}, k \in \mathcal{G},$ and $P_k=-P^0_{d_k}, k \in
\mathcal{L}.$

\textbf{Assumption 1:} There is a solution $\delta^*$ of equations
\eqref{eq.SEP} such that $|\delta_{kj}^*| \le \gamma < \pi/2$ for
all the transmission lines $\{k,j\} \in \mathcal{E}.$

We recall that for almost all power systems this assumption holds
true if we have the following synchronization condition, which is
established in \cite{Dorfler:2013},
\begin{align}
\label{eq.SynchronizationCondition}
||L^{\dag}p||_{\mathcal{E},\infty} \le \sin\gamma.
\end{align}
Here, $L^\dag$ is the pseudoinverse of the network Laplacian
matrix, $p=[P_1,...,P_{|\mathcal{N}|}]^T,$ and
$||x||_{\mathcal{E},\infty}=\max_{\{i,j\}\in
\mathcal{E}}|x(i)-x(j)|.$ In the sequel, we denote as
$\Delta(\gamma)$ the set of equilibrium points $\delta^*$
satisfying  that $|\delta_{kj}^*| \le \gamma<\pi/2, \forall \{k,j\}\in
\mathcal{E}.$ Then, any equilibrium point in this set is a stable operating point
\cite{Dorfler:2013}.

\subsection{Electronics-based Emergency Control Problem}
\label{sec.formulation}

In normal conditions, a power grid operates at a stable equilibrium point of
the pre-fault dynamics. After the initial disturbance (in this paper we consider the disturbance of a line tripping) 
the system evolves according to the fault-on dynamics laws and moves away
from the pre-fault equilibrium point $\delta^*_{pre}$. 
At the clearing time $\tau_{clearing}$, the fault is cleared, the system is at the
fault-cleared state $\delta_0=\delta_F(\tau_{clearing})$, and then the tripped line is reclosed. Hence, the system configuration is the same as pre-fault one and the power system experiences the post-fault transient dynamics. The
transient stability of the post-fault dynamics is certified if the post-fault
dynamics converges from the fault-cleared state to the post-fault
stable equilibrium point $\delta^*_{post}$, or more clearly, if the  fault-cleared state stays inside the region of attraction of the post-fault
stable equilibrium point.

In this paper, we assume that when a fault of line tripping happens, then the system operator can immediately send signals to synchronverters to simultaneously adjust the inertia and damping of the imitated generator without any communication and regulation delays (possible ways to deal with the issue of computation and regulation delays will be discussed in Section IV.C). We also assume that the tuned values of inertia and damping in these imitated generators can be kept in at least a time period $[0,\tau_{clearing}].$ Our emergency control problem is how to appropriately tune the inertia
and damping of the imitated generators to compensate for the disturbance such that after the given clearing time $\tau_{clearing},$ the fault-cleared state is still inside the region of attraction of the post-fault
stable equilibrium point $\delta^*_{post}$. 

If this objective can be obtained, then at the clearing time $\tau_{clearing},$ the fault is cleared, the inertia and damping of the imitated generators are brought back to their initial values, and the power system will evolve according to the post-fault dynamics from the fault-cleared state to the stable post-fault equilibrium point.

\section{Quadratic Lyapunov Function-based Transient Stability Certificate}
\label{sec.certificates}

In this section, we recall our recently introduced quadratic Lyapunov function-based transient stability certificate for power systems in \cite{VuTuritsyn:2015TAC}.  For this end, we
separate the nonlinear couplings and the linear terminal system in
\eqref{eq.structure-preserving}. For brevity, we denote the stable
post-fault equilibrium point for which we want to certify
stability as $\delta^*.$ Consider the state vector $x =
[x_1,x_2,x_3]^T,$ which is composed of the vector of generator's
angle deviations from equilibrium $x_1 = [\delta_1 -
\delta_1^*,\dots, \delta_{|\mathcal{G}|} -
\delta_{|\mathcal{G}|}^*]^T$, their angular velocities $x_2 =
[\dot\delta_1,\dots,\dot\delta_{|\mathcal{G}|}]^T$, and vector of
load buses' angle deviation from equilibrium
$x_3=[\delta_{{|\mathcal{G}|}+1}-\delta_{{|\mathcal{G}|}+1}^*,\dots,\delta_{|\mathcal{N}|}-\delta_{|\mathcal{N}|}^*]^T$.
Let $E$ be the incidence matrix of the  graph $\mathcal{G}(\mathcal{N},\mathcal{E})$, so
that $E[\delta_1,\dots,\delta_{|\mathcal{N}|}]^T =
[(\delta_k-\delta_j)_{\{k,j\}\in\mathcal{E}}]^T$. Let the matrix $C$ be
$E[I_{m\times m} \;O_{m \times n};O_{(n-m) \times 2m} \; I_{(n-m)\times (n-m)}].$ Then 
$$Cx=E[\delta_1-\delta_1^*,\dots,\delta_{|\mathcal{N}|}-\delta_{|\mathcal{N}|}^*]^T=[(\delta_{kj}-\delta_{kj}^*)_{\{k,j\}\in\mathcal{E}}]^T.$$
Consider the vector of nonlinear interactions $F$ in the simple trigonometric form: $
F(Cx)=[(\sin\delta_{kj}-\sin\delta^*_{kj})_{\{k,j\}\in\mathcal{E}}]^T.$ Denote
the matrices of moment of inertia, frequency controller  action on governor, and  frequency coefficient of load
as $M_1=\emph{\emph{diag}}(m_1,\dots,m_{|\mathcal{G}|}), D_1=\emph{\emph{diag}}(d_1,\dots,d_{|\mathcal{G}|})$ and $M=\emph{\emph{diag}}(m_1,\dots,m_{|\mathcal{G}|},d_{|\mathcal{G}|+1},\dots,d_{|\mathcal{N}|}).$

In state space representation, the power system \eqref{eq.structure-preserving} can be then expressed in the
following compact form:
\begin{align}
\dot{x}_1 &=x_2 \nonumber \\
\dot{x}_2 &=M_1^{-1}D_1x_2-S_1M^{-1}E^TSF(Cx)  \\
\dot{x}_3 &= -S_2M^{-1}E^TS F(Cx) \nonumber
\end{align}
where $S=\emph{\emph{diag}}(a_{kj})_{\{k,j\}\in \mathcal{E}},
S_1=[I_{m\times m}\quad O_{m\times n-m}], S_2=[O_{n-m\times m} \quad I_{n-m\times n-m}], n=|\mathcal{N}|, m=|\mathcal{G}|.$
Equivalently, we have
\begin{equation}\label{eq.Bilinear}
 \dot x = A x - B F(C x),
\end{equation}
with the matrices $A,B$ given by the following expression:
\begin{align*}
A=\left[
        \begin{array}{ccccc}
          O_{m \times m} \qquad & I_{m \times m}  \qquad & O_{m \times n-m}\\
          O_{m \times m} \qquad & -M_1^{-1}D_1 \qquad & O_{m \times n-m} \\
          O_{n-m \times m} \qquad &O_{n-m \times m} \qquad & O_{n-m \times n-m}
        \end{array}
      \right],
\end{align*}
and $$
 B= \left[
        \begin{array}{ccccc}
          O_{m \times |\mathcal{E}|}; \quad
          S_1M^{-1}E^TS; \quad
          S_2M^{-1}E^TS
        \end{array}
      \right].$$

The construction of quadratic Lyapunov function is based on the bounding of the nonlinear term $F$ by linear functions of the angular differences. Particularly, 
we observe that for all values of
$\delta_{kj} = \delta_k - \delta_j$  staying inside the polytope $\mathcal{P}$ defined by the inequalities $|\delta_{kj}|  \le\pi/2,$ we have:
\begin{align}
 g_{kj}(\delta_{kj}-\delta_{kj}^*)^2 \le (\delta_{kj}-\delta_{kj}^*)(\sin\delta_{kj} - \sin\delta_{kj}^*) \le (\delta_{kj}-\delta_{kj}^*)^2
\end{align}
where
\begin{align}
g_{kj}=\min \{\frac{1-\sin\delta_{kj}^*}{\pi/2-\delta_{kj}^*},
\frac{1+\sin\delta_{kj}^*}{\pi/2+\delta_{kj}^*}\} =
\frac{1-\sin|\delta_{kj}^*|}{\pi/2-|\delta_{kj}^*|}
\end{align}
Let $g=\min_{\{k,j\}\in \mathcal{E}}g_{kj}.$ Then, 
$g(\delta_{kj}-\delta_{kj}^*)^2 \le (\delta_{kj}-\delta_{kj}^*)(\sin\delta_{kj} - \sin\delta_{kj}^*) \le (\delta_{kj}-\delta_{kj}^*)^2.$

For each transmission line $\{k,j\}$ connecting generator buses $k$ and $j,$ define the corresponding flow-in boundary segment 
$\partial\mathcal{P}_{kj}^{in}$ of the polytope $\mathcal{P}$ by equations/inequations
$|\delta_{kj}|=\pi/2$ and $\delta_{kj}\dot{\delta}_{kj} < 0,$
and the flow-out boundary segment
$\partial\mathcal{P}_{kj}^{out}$  by
$|\delta_{kj}|=\pi/2$ and $\delta_{kj}\dot{\delta}_{kj} \ge 0.$
Consider the qudratic Lyapunov function $V(x)=x^TPx$ and define the following minimum value of the Lyapunov function $V(x)$ over the flow-out
boundary $\partial\mathcal{P}^{out}$ as:
\begin{align}\label{eq.Vmin}
 V_{\min}=\mathop {\min}\limits_{x \in \partial\mathcal{P}^{out}} V(x),
\end{align}
where $\partial\mathcal{P}^{out}$ is the union of
$\partial\mathcal{P}_{kj}^{out}$ over all the transmission lines $\{k,j\}\in \mathcal{E}$ connecting generator buses.  We have the following
result, which is a corollary of Theorem 1 in \cite{VuTuritsyn:2015TAC}. Hence, the proof is omitted.

\begin{theorem} (Transient Stability Certificate)

\label{thr.StabilityAssessment}
  \emph{Consider a power system with the post-fault equilibrium point $\delta^* \in \Delta(\gamma)$ and the fault-cleared state $x_0$ staying in the polytope $\mathcal{P}.$ Assume that there exists a positive definite matrix $P$ such that
  \begin{align}
\label{eq.LMI}
\left[%
\begin{array}{cc}
 \bar{A}^TP+P\bar{A} +  \dfrac{(1-g)^2}{4}C^TC     & PB \\
 B^TP  & -I \\
\end{array}%
\right] \le 0 
\end{align}
and 
\begin{align}
V(x_0) < V_{\min}
\end{align}
where $\bar{A}=A-\dfrac{1}{2}(1+g)BC.$
  Then, the system trajectory of \eqref{eq.structure-preserving} will converge from the fault-cleared state $x_0$ to the stable equilibrium point $\delta^*.$}
\end{theorem}

Therefore, a sufficient condition for the transient stability of the post-fault dynamics is the existence of a positive definite matrix $P$ satisfying the LMI \eqref{eq.LMI}
and the Lyapunov function at the fault-cleared state is small than the critical value $V_{\min}$ defined as in \eqref{eq.Vmin}. We will utilize this condition to design the emergency control in the next section.

\section{Synchronverter-based Emergency Control Design}
\label{sec.emergencycontrol}

\subsection{Control design}

In this section, we solve the problem of emergency control described in Section \ref{sec.formulation}, in which we maintain the power systems transient stability when a fault causes tripping of a line $\{u,v\}$. In particular, we will tune the inertia and damping of synchronverter-integrated generators such that at the fixed clearing time  $\tau_{clearing},$ the fault-cleared state $x_0$ is still inside the region of attraction of the 
post-fault equilibrium point $\delta^*.$ Assume that the tuned inertia and damping can be kept during the time period $[0,\tau_{cleating}].$ Applying Theorem 1, our objective is that: given a positive definite matrix $P$ satisfying the LMI \eqref{eq.LMI}, we will find the intertia and damping of the imitated generators such that the fault-cleared state $x_0$ satisfies $V(x_0)<V_{\min}.$

Indeed, the fault-on dynamics with the tuned inertia and damping is described by
\begin{align}
\label{eq.Faulton}
\dot{x}_F=A(m,d)x_F-B(m,d)F(Cx_F) +B(m,d) D_{\{u,v\}}\sin\delta_{F_{uv}},
\end{align}
where $D_{\{u,v\}}$ is the vector to extract the $\{u,v\}$ element
from the vector of nonlinear interactions $F,$ while $A(m,d)$ and $B(m,d)$ are the new system matrices $A,B$ obtained after the inertia and damping are tuned. 
Note that the system matrix $C$ is invariant to the changes of inertia and damping.

We have the following  center result of this paper:

\begin{theorem} (Emergency control design)

\label{thr.ECdesign} \emph{Assume that there exist a positive definite
matrix $P$ of size $(|\mathcal{N}|+|\mathcal{G}|)$ satisfying the LMI \eqref{eq.LMI}. Let $\mu=\dfrac{\tau_{clearing}}{V_{\min}}$ 
where $V_{\min}$ is defined as in \eqref{eq.Vmin}. If there exist inertia and damping of the imitated generators such that
\begin{align}
\label{eq.ECcondition2}
& \bar{A}(m,d)^TP+P\bar{A}(m,d) +  \dfrac{(1-g)^2}{4}C^TC  \nonumber \\& + PB(m,d)B(m,d)^TP \nonumber \\& +\mu PB(m,d)D_{\{u,v\}}D_{\{u,v\}}^TB(m,d)^TP  \le 0,
\end{align}
then,  the fault-cleared state $x_0=x_F(\tau_{clearing})$ resulted from the fault-on dynamics \eqref{eq.Faulton}  is still inside the
region of attraction of the post-fault equilibrium point $\delta^*$, and the
post-fault dynamics following the tripping and reclosing of the line $\{u,v\}$ will return to
the original stable operating condition.}
\end{theorem}

\emph{Proof:} See Appendix
\ref{appen.ECdesign}.

\begin{remark}
Note that the inequality \eqref{eq.ECcondition2} can be rewritten as 
\begin{align}
\label{eq.ECcondition3}
 &\bar{A}(m,d)^TP+P\bar{A}(m,d) +  \dfrac{(1-g)^2}{4}C^TC  \nonumber \\&+ P\bar{B}(m,d)\bar{B}(m,d)^TP  \le 0,
\end{align}
where $\bar{B}(m,d)=[B(m,d) \; \; \sqrt{\mu}B(m,d)D_{\{u,v\}}].$ By Schur complement, inequality \eqref{eq.ECcondition3} is equivalent with
\begin{align}
\label{eq.ECcondition4}
\left[%
\begin{array}{cc}
 \bar{A}(m,d)^TP+P\bar{A}(m,d) +  \dfrac{(1-g)^2}{4}C^TC     & P\bar{B}(m,d) \\
 \bar{B}(m,d)^TP  & -I \\
\end{array}%
\right] \le 0.
\end{align}
When the inertia of synchronverter-integrated generators are fixed, the damping of these generators enters linearly in the system matrices $A(m,d), B(m,d)$
and hence, the inequality \eqref{eq.ECcondition4} is an LMI of these variables. Also, when the damping is fixed, we obtain an LMI with variable as the inverse of inertia. In both these cases, we can quickly solve \eqref{eq.ECcondition4} by convex optimizations.
Therefore, the inequality \eqref{eq.ECcondition4} can be solved in polynomial time by a heuristic algorithm, in which we fix the inertia or damping and then use the convex optimizations to solve the corresponding LMI to obtain the optimum value of the remaining variable.
\end{remark}

\subsection{Procedure for emergency control}
\label{sec.procedure}

Given the power system under a line tripping and the clearing time $\tau_{clearing},$ we have the following procedure to tune the inertia and damping in the 
synchronverter-based emergency control:

\begin{itemize}
\item [1)] Find a positive definite matrix $P$ satisfying the LMI \eqref{eq.LMI}.
\item [2)] Calculate the minimum value $V_{\min}$ defined as in \eqref{eq.Vmin}.
\item [3)] Let $\mu=\dfrac{\tau_{clearing}}{V_{\min}}.$
\item [4)] Find the inertia and damping of the synchronverter-integrated generators such that the inequality \eqref{eq.ECcondition4} is satisfied. One approach to quickly come up with the solution is to use the heuristic algorithm described in the previous section.
\item [5)] If there is no such inertia and damping, then repeat from step 1).
\item [6)] If such values of inertia and damping exist, then the synchronverters will be used to tune the inertia and damping of the imitated generators and keep these values
during the time period $[0,\tau_{clearing}].$ At the clearing time $\tau_{clearing},$ the fault is cleared and the inertia and damping of the imitated generators are tuned back to their initial values.
\end{itemize}

\subsection{Discussions on computation and regulation delays}

Computation and regulation delays may make the proposed emergency control scheme in this paper not yet ready for industrial deployment. To overcome this obstacle, we propose
the following off-line computational tasks:

\begin{itemize}
\item [1)] For each line tripping $\{u,v\},$ we off-line check if there exists a positive definite matrix $P$ satisfying the LMI \eqref{eq.LMI} and the inequality \eqref{eq.ECcondition4}
where $A(m,d)=A, B(m,d)=B.$ If such matrix exists, then the line tripping is safe without any emergency control. The LMI \eqref{eq.LMI} is only dependent on the system matrices $A,B,C,$ and thus is checkable before hand. The inequality \eqref{eq.ECcondition4} is also dependent on $V_{\min}$ and thus on the equilibrium point. However, we observe that the equilibrium point in practice usually stays inside a small region. As such, we can obtain some lower bound for $V_{\min}$ for all of these equilibrium points. By this way, we can check the inequality \eqref{eq.ECcondition4} for a wide range of post-fault equilibrium points. We believe that by this way we can certify that most of the line trippings in practice are safe.
\item [2)] For the remaining unsafe line trippings, we calculate before hand if there exist optimum values of inertia and damping of the imitated generators such that the inequality \eqref{eq.ECcondition4} is satisfied. Again, we need to use some bound for $V_{\min}$ as above. If these values exist, then we can apply the emergency control described in the previous section right after the fault happens.
\item [3)] For some unsafe line tripping that there are no inertia and damping satisfying the inequality \eqref{eq.ECcondition4}, we may apply traditional emergency control schemes such as load shedding to quickly stabilize the system.
\end{itemize}

Another way is to allow for a time period of $[0,\tau_{delay}]$ to compensate for the time of computation and regulation. In this period, the fault-on trajectory evolves according to fault-on dynamics \eqref{eq.Faulton} where $A(m,d)=A, B(m,d)=B.$ Again, by finding the positive definite matrix $P$ satisfying both the LMI \eqref{eq.LMI} and the inequality \eqref{eq.ECcondition4} where $A(m,d)=A, B(m,d)=B,$ we can bound the fault-on dynamics during the delayed period $[0,\tau_{delay}].$ This bound will help us to design the inertia and damping in the period $[\tau_{delay},\tau_{clearing}]$ such that the fault-cleared state at the clearing time $\tau_{clearing}$ is still inside the region of attraction of the post-fault equilibrium point.

\subsection{Other way to design the emergency control}

This section presents another way to determine the optimum value of the inertia and damping of the imitated generators such that the fault-cleared state satisfies $V(x_0)<V_{\min}$ 
whwere $V(x)=x^TPx$ and $V_{\min}$ is defined as in \eqref{eq.Vmin}. The reason is that there are possible situations where we fail to find the inertia and damping by the procedure described in Section \ref{sec.procedure} because we require a common Lyapunov function $P$ for both fault-on dynamics and post-fault dynamics. Our observation is that it is easier to find diiferent Lyapunov functions for different dynamics. Therefore, we propose the following procedure to find inertia and damping:

\begin{itemize}
\item [1)] Find a positive definite matrix $P$ satisfying the LMI \eqref{eq.LMI}.
\item [2)] Calculate the minimum value $V_{\min}$ defined as in \eqref{eq.Vmin}.
\item [3)] Let $\mu=\dfrac{\tau_{clearing}}{V_{\min}}.$
\item [4)] Vary the inertia and damping of the synchronverter-integrated generators, and for each fixed value of inertia and damping, find a positive definite matrix $\tilde{P}=P(m,d)$ such that 
\begin{align}
\label{eq.ECcondition5}
\left[%
\begin{array}{cc}
 \bar{A}(m,d)^T\tilde{P}+\tilde{P}\bar{A}(m,d) +  \dfrac{(1-g)^2}{4}C^TC     & \tilde{P}\bar{B}(m,d) \\
 \bar{B}(m,d)^T\tilde{P}  & -I \\
\end{array}%
\right] \le 0,
\end{align}
and 
\begin{align}
\label{eq.AdditionalCondition}
P \le P(m,d).
\end{align}

\item [5)] If there is no such matrix $P(m,d)$, then repeat from step 1) or setp 4) with new values of internia and damping.
\item [6)] If such values of inertia and damping exist, then the synchronverters will be used to tune the inertia and damping of the imitated generators and keep these values
during the time period $[0,\tau_{clearing}].$ At the clearing time $\tau_{clearing},$ the fault is cleared and the inertia and damping of the imitated generators are tuned back to their initial values.
\end{itemize}

Since the matrix $P(m,d)$ satisfies the LMI \eqref{eq.ECcondition5}, similar to Theorem 2 we can prove that $x_0^TP(m,d)x_0 < V_{\min}.$ This together with \eqref{eq.AdditionalCondition} leads to $x_0^TPx_0 \le x_0^TP(m,d)x_0 <V_{\min}.$ Applying Theorem 1, we conclude that the fault-cleared state stays inside the region of attraction and therefore the post-fault dynamics is stable.

\section{Numerical Validation}
\label{sec.simulations}

For illustrating the concept of this paper, we consider the simple yet non-trivial system of three generators, one of which is the renewable generator (generator 1) integrated with the synchronverter. 
The susceptance of the transmission lines are assumed at fixed values $B_{12}=0.739$ p.u., $B_{13}=1.0958$ p.u., and
$B_{23}=1.245$ p.u. Also, the inertia and damping of all the conventional and imitated generator at the normal working condition are $m_k=2$ p.u., $d_k=1$ p.u. Assume that the line between generators 1 and 2 is tripped, and then reclosed at the clearing time $\tau_{clearing}=200ms,$ and during the fault-on dynamic stage the time-invariant terminal voltages and mechanical torques given in Tab. \ref{tab.data}.

\begin{table}[ht!]
\centering
\begin{tabular}{|c|c|c|}
  \hline
  Node & V (p.u.) & P (p.u.) \\
  \hline
  1 & 1.0566 & -0.2464 \\
  2 & 1.0502 & 0.2086 \\
  3 & 1.0170 & 0.0378 \\
  \hline
\end{tabular}
\caption{Voltage and mechanical input} \label{tab.data}
\end{table}
The pre-fault and post-fault equilibrium point is calculated from \eqref{eq.SEP}: $\delta^*=[-0.6634\;
   -0.5046\;
   -0.5640 \;0\;0\;0]^T.$ Hence, the equilibrium point stays in the polytope defined by the inequality $|\delta_{kj}|<\pi/10.$ As such, we can take 
   $g=(1-\sin(\pi/10))/(\pi/2-\pi/10).$ Using CVX in MATLAB to solve the LMI \eqref{eq.LMI}, we can obtain the Lyapunov function $V(x)=x^TPx$ where
   \begin{align}
   \label{eq.LyaFunction}
   P=\left[%
\begin{array}{cccccc}
   2.8401  &  1.9098  &  1.9812  &  4.5726   & 4.4349  &  4.4563\\
    1.9098  &  2.7949 &   2.0263  &  4.4393  &  4.5628  &  4.4578\\
    1.9812  &  2.0263  &  2.7235  &  4.4502  &  4.4644   & 4.5480\\
    4.5726  &  4.4393  &  4.4502  & 18.4333  & 17.5302  & 17.6662\\
    4.4349  &  4.5628  &  4.4644  & 17.5302  & 18.3632   &17.7364\\
    4.4563  &  4.4578  &  4.5480  & 17.6662  & 17.7364  & 18.2271\\
\end{array}%
\right]
   \end{align}
Then the minimum value $V_{\min}$ is $V_{\min}=0.8139$ and thus $\mu=\tau_{clearing}/V_{\min}=0.2457.$ 

\subsection{Procedure IV.B}
\begin{figure}[t!]
\centering
\includegraphics[width = 3.2in]{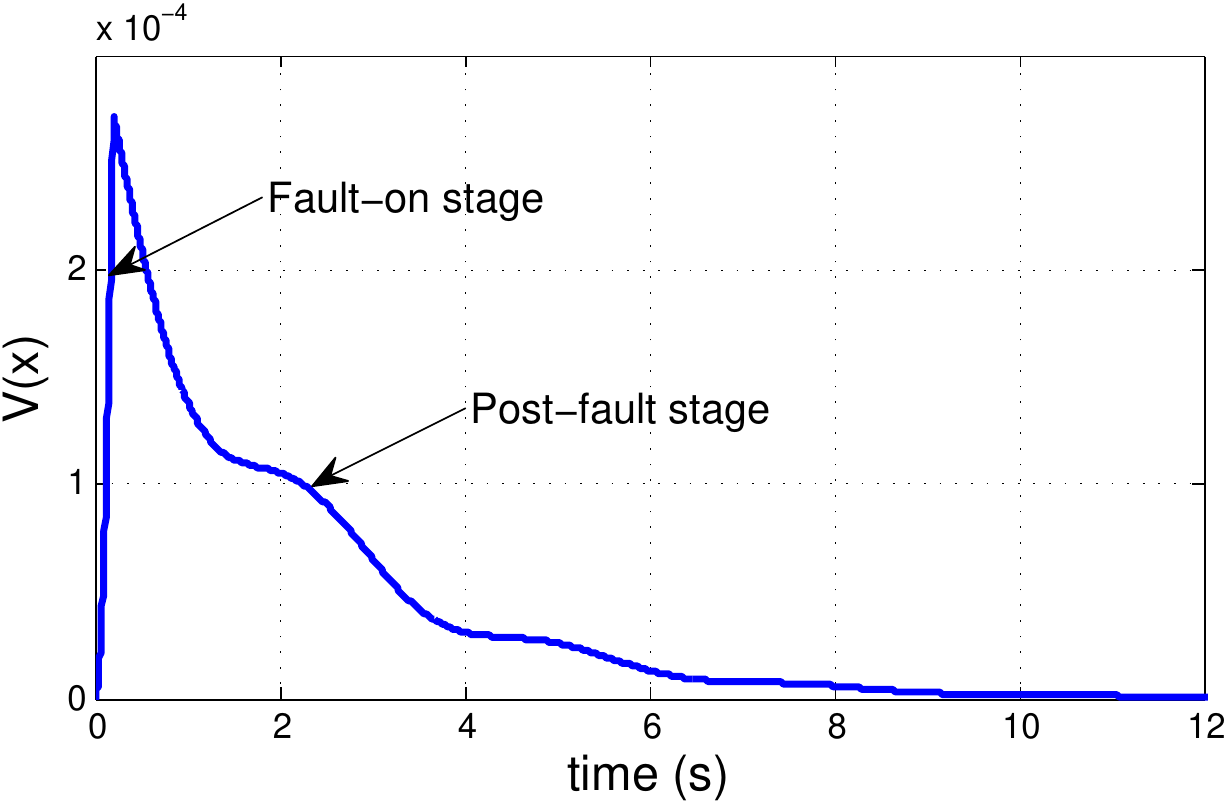}
\caption{Variations of the quadratic Lyapunov function $V(x)=x^TPx=(\delta-\delta^*)^TP(\delta-\delta^*)$ during the fault-on and post-fault dynamics.}
\label{fig.EmergencyControl_Lyapunov}
\end{figure}

\begin{figure}[t!]
\centering
\includegraphics[width = 3.2in]{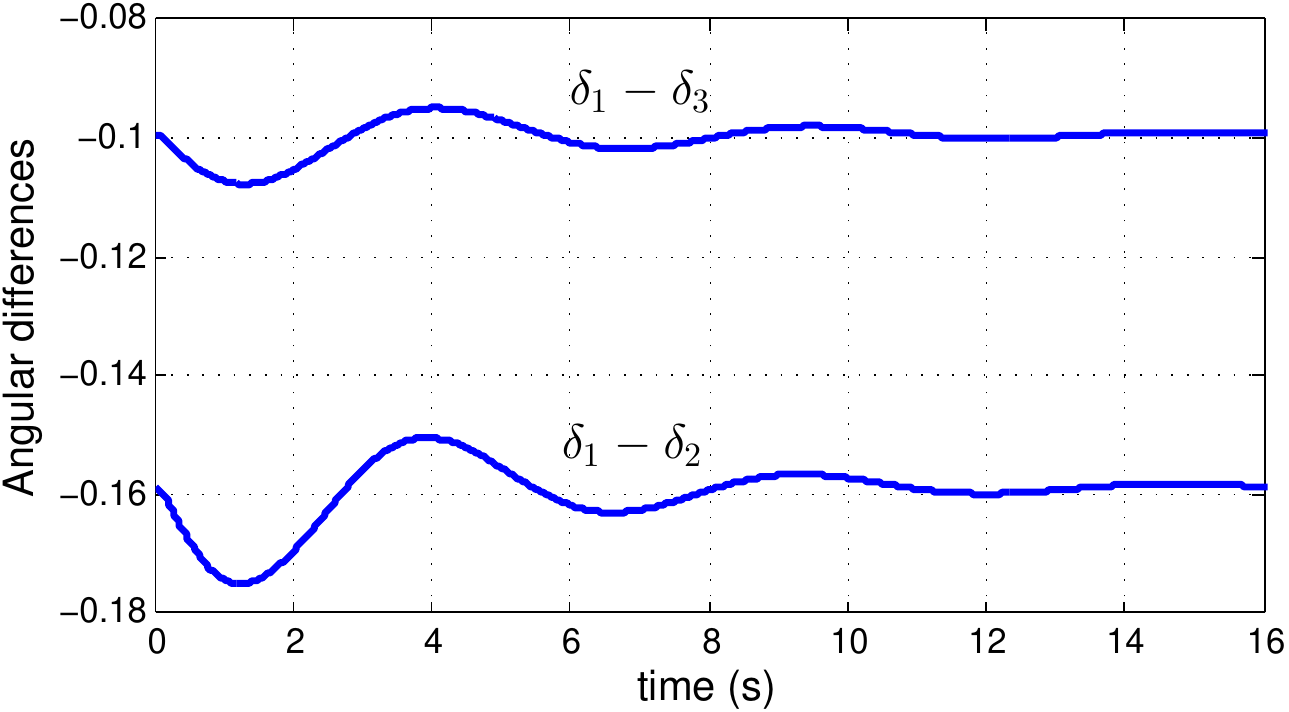}
\caption{Variations of the angular differences during the fault-on and post-fault dynamics.}
\label{fig.EmergencyControl_Angular}
\end{figure}

Again, using the CVX to solve the LMI \eqref{eq.ECcondition4}
where the inertia of the imitated generator is fixed, we obtain the optimum damping $d_1=1$ p.u. Figure \ref{fig.EmergencyControl_Lyapunov} shows that the Lyapunov function $V(x)=x^TPx=(\delta-\delta^*)^TP(\delta-\delta^*)$ evolves from $0$ to some value during the fault-on dynamics and then converge to $0$ during the post-fault dynamics. This means that the post-fault power system is transiently stable under the effect of the proposed emergency control. Similarly, Fig. \ref{fig.EmergencyControl_Angular} shows that the angular differences
of the generators (i.e. $\delta_1-\delta_2$ and $\delta_1-\delta_3$) converge to the constant values during the post-fault dynamics, confirming the stability of the system.

\subsection{Procedure IV.D}
\begin{figure}[t!]
\centering
\includegraphics[width = 3.2in]{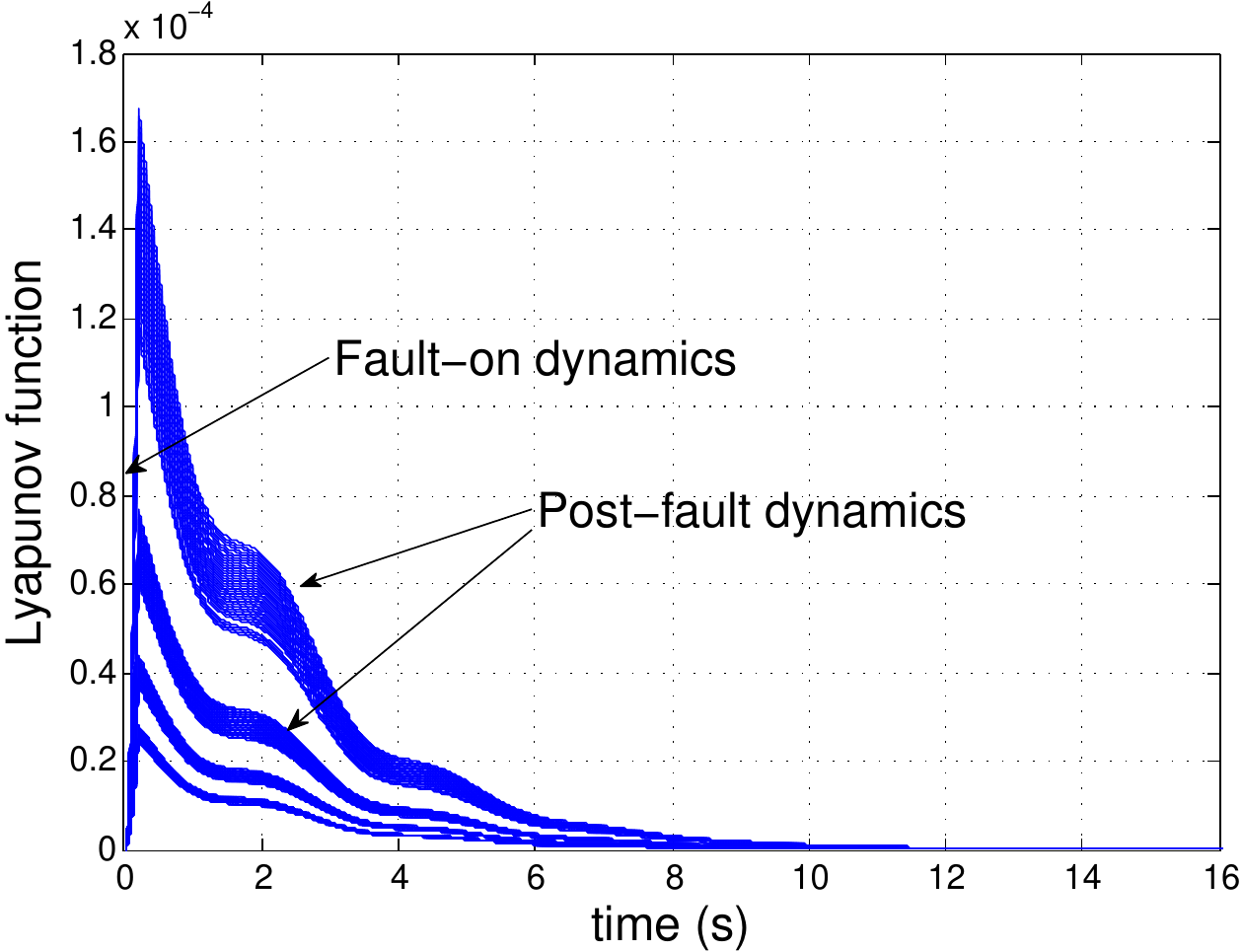}
\caption{Variations of the quadratic Lyapunov function $V(x)=x^TPx=(\delta-\delta^*)^TP(\delta-\delta^*)$ during the post-fault and fault-on dynamics with different values of inertia and damping.}
\label{fig.EmergencyControlSet_Lyapunov}
\end{figure}

This section illustrates the effectiveness of the procedure described in Section IV.D. We use the Lyapunov function $V(x)=x^TPx$, where $P$ is given in \eqref{eq.LyaFunction},
for the post-fault dynamics, and then solve the LMI \eqref{eq.ECcondition5}-\eqref{eq.AdditionalCondition} to find the Lyapunov function $x^TP(m,d)x$ for the fult-on dynamics for each varied value of new inertia and damping. We can see that there are many values of inertia and damping such that the LMI \eqref{eq.ECcondition5}-\eqref{eq.AdditionalCondition} has positive definite solution. This means that there are many values of inertia and damping to compensate for the dynamics deviation caused by the faulted line ans thereby stabilize the power systems. We draw in Fig. \ref{fig.EmergencyControlSet_Lyapunov} the dynamics of Lyapunov function $V(x)=x^TPx$ during the fault-on and post-fault dynamics, where the fault-on dynamics is  controlled by different feasible values of inertia and damping. This confirms that there are many suitable values of inertia and damping for the emergency control.

\section{Conclusions and path forward}
\label{sec.conclusion}

This paper was dedicated to bring into the attention of the control community the problem of emergency control in power systems. In particular, we discussed a novel emergency control for power grids with high penetration of renewables by exploiting the emerging power electronics resources. For this end, we modeled the structure preserving model of such systems where the renewables are integrated with the synchronverters which will control the renewable generator to mimic the dynamics of the conventional generators but with tunable damping and inertia. On top of this model, we formulated the emergency control to maintain the transient stability of post-fault dynamics following a given line tripping by intelligently adjusting the damping and inertia of the imitated generators. Applying our recently introduced quadratic Lyapunov function-based transient stability certificate, we showed that this problem can be solved through a number of convex optimizations in the form of linear matrix inequalities. The numerical simulations showed that this emergency control is effective to recover transient stability after critical line trippings.

We also discussed possible ways to make this novel emergency scheme ready for industrial employment. Particularly, we sketched ways to address the issues of computation and regulations delays, either by offline scanning contingencies and calculating the emergency actions before hand, or by allowing specific delayed time for computation. Future works would demonstrate the proposed emergency control scheme on large IEEE prototype and large dynamic realistic power systems with renewable generation at various locations and with different levels of renewable penetration. Also, we  would investigate the potential values of transmission facilities, which are ubiquitously equipped in the existing power grids such as FACTS devices, to intelligently control the transmission network as an alternative remedial action to emergency situations.

\section{Appendix}

\subsection{Proof of Theorem \ref{thr.ECdesign}}
\label{appen.ECdesign}

We have the derivative of $V(x)$ along the fault-on trajectory \eqref{eq.Faulton} as follows:
\begin{align}
\dot{V}(x_F) &= \dot{x_F}^TPx_F + x_F^TP\dot{x_F}  = x_F^T(A(m,d)^TP+PA(m,d))x_F\nonumber\\& -2 x_F^TPB(m,d)F + 2x_F^T PB(m,d)D_{\{u,v\}}\sin\delta_{F_{uv}} \nonumber \\
&=W(x_F) -S^TS  + 2x_F^T PB(m,d)D_{uv}\sin\delta_{F_{uv}} \nonumber \\
& +x_F^T\big[A(m,d)^TP+PA(m,d)- C^TK_1^TK_2C + R^TR \big]x_F, 
\end{align}
where
\begin{align}
W(x_F)&=(F-gCx_F)^T(F-Cx_F) \nonumber\\
R&=B(m,d)^TP-\frac{1}{2}(1+g)C \nonumber\\
S&=F+(B(m,d)^TP-\frac{1}{2}(1+g)C)x_F. \nonumber
\end{align}

On the other hand
\begin{align}
&2x_F^T PB(m,d)D_{\{u,v\}}\sin\delta_{F_{uv}} \le  \dfrac{1}{\mu}\sin^2\delta_{F_{uv}} +\nonumber\\& 
\mu x_F^T PB(m,d)D_{\{u,v\}}D_{\{u,v\}}^TB(m,d)^TP x_F
\end{align}
Therefore,

\begin{align}
\dot{V}(x_F)\le W(x_F) -S^TS +x_F^T\tilde{Q}x_F +\frac{1}{\mu}\sin^2\delta_{F_{uv}} 
\end{align}
where $\tilde{Q}=A(m,d)^TP+PA(m,d)- C^TK_1^TK_2C + R^TR +\mu PB(m,d)D_{\{u,v\}}D_{\{u,v\}}^TB(m,d)^TP$. Note that $W(x_F)\le 0, \forall x_F\in \mathcal{P},$ and
\begin{align}
\tilde{Q}&=\bar{A}(m,d)^TP+P\bar{A}(m,d) +  \dfrac{(1-g)^2}{4}C^TC \nonumber \\&   + PB(m,d)B(m,d)^TP \nonumber \\& +\mu PB(m,d)D_{\{u,v\}}D_{\{u,v\}}^TB(m,d)^TP \le 0.
\end{align}
Therefore,
\begin{align}
\label{eq.Vfault}
\dot{V}(x_F) \le \frac{1}{\mu}\sin^2\delta_{F_{uv}}\le \frac{1}{\mu}, 
\end{align}
whener $x_F$ in the polytope $\mathcal{P}.$

We will prove that the
fault-cleared state $x_F(\tau_{clearing})$ is in the set
$\mathcal{R}=\{x\in \mathcal{P}:V(x)<V_{\min}\},$ which, from Theorem 1, is a subset of the region of attraction of the stable post-fault equilibrium point. 

Note that the boundary of $\mathcal{R}$ is constituted of the segments on flow-in boundary $\partial\mathcal{P}^{in}$ and the segments on the sublevel sets 
of the Lyapunov function. It is easy to see that the
flow-in boundary $\partial\mathcal{P}^{in}$ 
prevents the fault-on dynamics \eqref{eq.Faulton} from escaping $\mathcal{R}.$
Assume that $x_F(\tau_{clearing})$ is not in the set
$\mathcal{R}.$ Then the fault-on trajectory can only escape
$\mathcal{R}$ through the segments which belong to sublevel set of
the Lyapunov function $V(x).$ Denote $\tau$ be the first time at
which the fault-on trajectory meets one of the boundary segments
which belong to sublevel set of the Lyapunov function $V(x).$
Hence $x_F(t) \in \mathcal{R}$ for all $0 \le t \le \tau.$ From
\eqref{eq.Vfault} and the fact that
$\mathcal{R}\subset \mathcal{P},$ we have

\begin{align}
\label{eq.contradiction}
V(x_F(\tau))-V(x_F(0)) &= \int_0^{\tau} \dot{V}(x_F(t))dt \le
\frac{\tau}{\mu} \nonumber \\&
< \frac{\tau_{clearing}}{\mu}=V_{\min}
\end{align}
Note that $x_F(0)$ is the pre-fault equilibrium point, and thus equals to post-fault equilibrium point.
Hence $V(x_F(0))=0.$ By definition, we have
$V(x_F(\tau))=V_{\min},$  which is a
contradiction with \eqref{eq.contradiction}.  $\square$

\section{Acknowledgements}
This work was partially supported by NSF, MIT/Skoltech, Masdar
initiatives and Ministry of Education and Science of Russian
Federation, Grant Agreement no. 14.615.21.0001.

\bibliographystyle{IEEEtran}
\bibliography{lff}

\newcommand{\noopsort}[1]{} \newcommand{\printfirst}[2]{#1}
  \newcommand{\singleletter}[1]{#1} \newcommand{\switchargs}[2]{#2#1}
\begin{thebibliography}{10}
\providecommand{\url}[1]{#1}
\csname url@samestyle\endcsname
\providecommand{\newblock}{\relax}
\providecommand{\bibinfo}[2]{#2}
\providecommand{\BIBentrySTDinterwordspacing}{\spaceskip=0pt\relax}
\providecommand{\BIBentryALTinterwordstretchfactor}{4}
\providecommand{\BIBentryALTinterwordspacing}{\spaceskip=\fontdimen2\font plus
\BIBentryALTinterwordstretchfactor\fontdimen3\font minus
  \fontdimen4\font\relax}
\providecommand{\BIBforeignlanguage}[2]{{%
\expandafter\ifx\csname l@#1\endcsname\relax
\typeout{** WARNING: IEEEtran.bst: No hyphenation pattern has been}%
\typeout{** loaded for the language `#1'. Using the pattern for}%
\typeout{** the default language instead.}%
\else
\language=\csname l@#1\endcsname
\fi
#2}}
\providecommand{\BIBdecl}{\relax}
\BIBdecl

\bibitem{119276}
C.~Lu and M.~Unum, ``Interactive simulation of branch outages with remedial
  action on a personal computer for the study of security analysis [of power
  systems],'' \emph{Power Systems, IEEE Transactions on}, vol.~6, no.~3, pp.
  1266--1271, Aug 1991.

\bibitem{6965423}
A.~Shrestha, V.~Cecchi, and R.~Cox, ``Dynamic remedial action scheme using
  online transient stability analysis,'' in \emph{North American Power
  Symposium (NAPS), 2014}, Sept 2014, pp. 1--6.

\bibitem{6939093}
J.~Mitra, ``Real-time remedial action screening using direct stability analysis
  methods,'' in \emph{PES General Meeting | Conference Exposition, 2014 IEEE},
  July 2014, pp. 1--1.

\bibitem{SPS}
P.~Anderson and B.~LeReverend, ``Industry experience with special protection
  schemes,'' \emph{Power Systems, IEEE Transactions on}, vol.~11, no.~3, pp.
  1166--1179, Aug 1996.

\bibitem{982194}
W.~Fu, S.~Zhao, J.~McCalley, V.~Vittal, and N.~Abi-Samra, ``Risk assessment for
  special protection systems,'' \emph{Power Systems, IEEE Transactions on},
  vol.~17, no.~1, pp. 63--72, Feb 2002.

\bibitem{141798}
S.~Nirenberg, D.~McInnis, and K.~Sparks, ``Fast acting load shedding,''
  \emph{Power Systems, IEEE Transactions on}, vol.~7, no.~2, pp. 873--877, May
  1992.

\bibitem{1461637}
A.~Zin, H.~Hafiz, and M.~Aziz, ``A review of under-frequency load shedding
  scheme on tnb system,'' in \emph{Power and Energy Conference, 2004. PECon
  2004. Proceedings. National}, Nov 2004, pp. 170--174.

\bibitem{6345547}
K.~Mollah, M.~Bahadornejad, N.-K. Nair, and G.~Ancell, ``Automatic
  under-voltage load shedding: A systematic review,'' in \emph{Power and Energy
  Society General Meeting, 2012 IEEE}, July 2012, pp. 1--7.

\bibitem{Koch2010}
S.~Koch, S.~Chatzivasileiadis, M.~Vrakopoulou, and G.~Andersson, ``Mitigation
  of cascading failures by real-time controlled islanding and graceful load
  shedding,'' \emph{Bulk Power System Dynamics and Control (iREP)-VIII, iREP
  Symposium}, pp. 1--19, 2010.

\bibitem{Zhong:2011}
Q.-C. Zhong and G.~Weiss, ``Synchronverters: Inverters that mimic synchronous
  generators,'' \emph{Industrial Electronics, IEEE Transactions on}, vol.~58,
  no.~4, pp. 1259--1267, April 2011.

\bibitem{VuTuritsyn:2015TAC}
T.~L. Vu and K.~Turitsyn, ``{A Framework for Robust Assessment of Power Grid
  Stability and Resiliency},'' \emph{Automatic Control, IEEE Trans.}, 2015, in
  revision, available: arXiv:1504.04684.

\bibitem{bergen1981structure}
A.~R. Bergen and D.~J. Hill, ``A structure preserving model for power system
  stability analysis,'' \emph{Power Apparatus and Systems, IEEE Transactions
  on}, no.~1, pp. 25--35, 1981.

\bibitem{Dorfler:2013}
F.~Dorfler, M.~Chertkov, and F.~Bullo, ``{Synchronization in complex oscillator
  networks and smart grids},'' \emph{Proceedings of the National Academy of
  Sciences}, vol. 110, no.~6, pp. 2005--2010, 2013.

\end{thebibliography}
\end{document}